\documentclass[a4paper]{jpconf}
\usepackage{graphicx}
\newcommand {\apgt} {\ {\raise-.5ex\hbox{$\buildrel>\over\sim$}}\ }
\newcommand {\aplt} {\ {\raise-.5ex\hbox{$\buildrel<\over\sim$}}\ }

\begin{document}
\title{$\gamma$-ray emission region located in the parsec scale jet of OJ287}

\author{I. Agudo$^{1,2}$, S. G.~Jorstad$^{2,3}$, A. P.~Marscher$^{2}$, V. M.~Larionov$^{3,4}$, J. L.~G\'omez$^{1}$, A. L\"{a}hteenm\"{a}ki$^{5}$, M. Gurwell$^{6}$, P. S. Smith$^{7}$, H. Wiesemeyer$^{8,9}$, C. Thum$^{10}$ and J. Heidt$^{11}$}

\address{$^{1}$Instituto de Astrof\'{i}sica de Andaluc\'{i}a, CSIC, Granada, Spain}
\address{$^{2}$Institute for Astrophysical Research, Boston University,Boston, USA}
\address{$^{3}$Astronomical Institute, St. Petersburg State University, St. Petersburg, Russia}
\address{$^{4}$Isaac Newton Institute of Chile, St. Petersburg Branch, St. Petersburg, Russia}
\address{$^{5}$Aalto University Mets\"{a}hovi Radio Observatory, Kylm\"{a}l\"{a}, Finland}
\address{$^{6}$Harvard--Smithsonian Center for Astrophysics, Cambridge, USA}
\address{$^{7}$Steward Observatory, University of Arizona, Tucson, USA}
\address{$^{8}$Instituto de Radio Astronom\'{i}a Milim\'{e}trica, Granada, Spain}
\address{$^{9}$Max-Plack-Institut f\"ur Radioastronomie, Bonn, Germany} 
\address{$^{10}$Institut de Radio Astronomie Millim\'{e}trique, St. Martin d'H\`{e}res, France} 
\address{$^{11}$ZAH, Landessternwarte Heidelberg, K\"{o}nigstuhl, Heidelberg, Germany}

\ead{iagudo@iaa.es}

\begin{abstract}
We report on the location of the $\gamma$-ray emission region in flares of the BL Lacertae object OJ287 at $>14$\,pc from the central supermassive black hole. 
We employ data from multi-spectral range (total flux and linear polarization) monitoring programs combined with sequences of ultra-high-resolution 7\,mm VLBA images. 
The correlation between the brightest $\gamma$-ray and mm flares is found to be statistically significant.
The two $\gamma$-ray peaks, detected by {\emph Fermi}-LAT, that we report here happened at the rising phase of two exceptionally bright mm flares accompanied by sharp linear polarization peaks.
The VLBA images show that these mm flares in total flux and polarization degree occurred in a jet region at $>14$\,pc from the innermost jet region.
The time coincidence of the brighter  $\gamma$-ray flare and its corresponding mm linear polarization peak evidences that both the $\gamma$-ray and mm outbursts occur $>14$\,pc from the central black hole.
We find two sharp optical flares occurring at the peak times of the two reported $\gamma$-ray flares.
This is interpreted as the $\gamma$-ray flares being produced by synchrotron self-Compton scattering of optical photons from the flares triggered by the interaction of moving knots with a stationary conical shock in the jet. 
\end{abstract}

\section{Introduction}
The hypothesis of the location of the $\gamma$--ray flare emission in blazars close to the mm jet cores ($>>1$\,pc downstream the supermassive black hole) started to have support from the observations made by the Energetic Gamma Ray Experiment Telescope (EGRET), aboard the Compton Gamma Ray Observatory.
The EGRET data suggested the connection of $\gamma$--ray flares with their mm wavelength counterpart \cite{Lahteenmaki:2003p5657}, the connection of $\gamma$--ray flares with superluminal ejection of blobs in the jets of some blazars  \cite{2001ApJ...556..738J}, and the relation of the flux density of the mm wavelength VLBI core with the $\gamma$--ray flux \cite{Jorstad:2001p5655}.

The unprecedented sensitivity of the Large Area Telescope (LAT) aboard the \emph{Fermi} Gamma Ray Space Telescope provides detailed $\gamma$--ray light curves of tens of bright blazars that, together with those from multi--spectral--range observing programs, has allowed to identify abrupt $\gamma$--ray flares with those at other spectral ranges and to provide robust evidence of the location of the $\gamma$--ray flare emitting region $>>1$\,pc from the central black hole; e.g. the cases of AO~0235+164 \cite{Agudo:2011p15946}, PKS~1510$-$089 \cite{Jorstad:2010p11830}, 3C~279 \cite{Abdo:2010p11811}, and 3C~454.3 \cite{Marscher:2010p11374}.
However, the location of the $\gamma$--ray emission --which is a critical issue to identify the $\gamma$-ray emission mechanism-- is still under debate for some blazars, as was raised in this conference, see Poutanen et al. (these proceedings) and Becerra et al. (these proceedings).

The technique used by \cite{Marscher:2010p11374}, \cite{Jorstad:2010p11830} and \cite{Agudo:2011p15946} is based on the use of ultra--high--resolution VLBI monitoring to resolve the innermost jet regions and to look for the correlation of their polarized emission with shorter wavelength observations along the spectrum. 

In this paper, where we extend on previous work \cite{Agudo:2011p14707}, we use this technique to investigate the location of the flaring $\gamma$--ray emission in the BL~Lacertae (BL~Lac) object {OJ287} ($z=0.306$), by applying, for the first time in this kind of studies, a Monte--Carlo--based method to evaluate the significance of the correlation with the \emph{Fermi}--LAT $\gamma$--ray light curve.

\section{Observations}
\begin{figure}
   \centering
   \includegraphics[clip,width=12.cm]{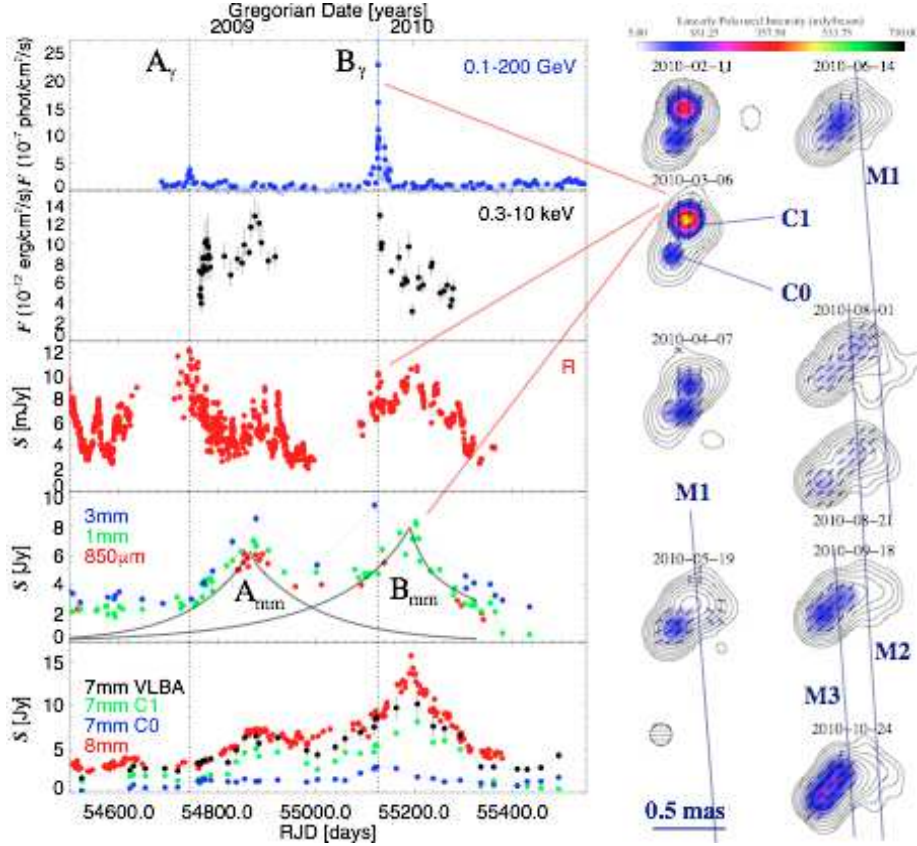}
   \caption{{\emph Left:} Light curves of OJ287 from $\gamma$-rays (top), to mm wavelengths (bottom). RJD = Julian Date $- 2400000.0$. {\emph Right:} Sequence of 7\,mm VLBA images from February to October 2010.
   Images were convolved with a $\rm{FWHM}=0.15$\,mas circular Gaussian beam.
   Contour levels start at 0.2 of the peak total intensity (6.32\,Jy/beam) and increment by factors of two. 
   The color scale represents linear polarization intensity, and black segments symbolize the distribution of position angles of the linear polarization. A movie of the total intensity time evolution corresponding to the same time span can be found here: {\tt https://w3.iaa.es/$\sim$iagudo/research/OJ287/OJ287FebOct2010.mov} .}
   \label{fig1}
\end{figure}

The photo-polarimetric observations presented here include 7\,mm VLBA maps at 0.15\,mas resolution (Fig.~\ref{fig1}), 3\,mm observations with the IRAM 30\,m Telescope, and optical observations (Figs.~\ref{fig1} and \ref{fig2}) that include data from the Calar Alto (2.2\,m Telescope), Steward (2.3 and 1.54\,m telescopes), Lowell (1.83\,m Perkins Telescope), St. Petersburg State University (0.4\,m Telescope), and Crimean Astrophysical  (0.7\,m Telescope) observatories, as well as photo-polarimetric data from \cite{Villforth:2010p11557}.
Our total flux light curves (Fig.~\ref{fig1}) come from the \emph{Fermi}-LAT $\gamma$-ray (0.1--200\,GeV) and \emph{Swift} X-ray (0.3--10\,keV) archives, and from the one by the Yale University SMARTS program.
To these, we add data from the SMA at 1.3\,mm and 850\,$\mu$m, from the IRAM 30\,m Telescope at 1.3\,mm, and from the Mets\"{a}hovi Radio Observatory 14\,m Telescope at 8\,mm.

The reduction of our data followed the procedures explained in \cite{Agudo:2011p14707}. 
In particular we followed \cite{Jorstad:2005p264} for the VLBA data, \cite{Jorstad:2010p11830} for the optical polarimetric data, \cite{Agudo:2006p203,Agudo:2010p12104} for the IRAM data,  \cite{Jorstad:2010p11830} for the \emph{Swift} data, and \cite{Marscher:2010p11374} for the \emph{Fermi} LAT data. 
However, for the $\gamma$-ray light curve in Fig.~\ref{fig1}, which covers a longer time range than the one presented in our previous work on OJ287 \cite{Agudo:2011p14707}, release v9r18p6 of the {\emph Fermi} Science Tools was used, instead of previous release (v9r15p2).
Release v9r18p6 produced significantly higher SNR results, and hence larger number of detections.
This allowed us to use here a time binning of 5 days during quiescent periods, and even 1 day or better around the two $\gamma$-ray peaks discussed below.  

\section{Results}

\subsection{1\,mm and 7\,mm flares located at the core}

We modeled the brightness distribution of the 7mm VLBA images of the source with a small number of circular Gaussian components.
The two brightest ones are Q0, that we identify here as the \emph{core} (i.e., the innermost jet region observable in our 7\,mm images.) and Q1, see Fig.~\ref{fig1}.
The identification of C0 as the innermost jet feature is justified by the decrease of the emission to the west of C1, and by the detection of superluminal motion of features M1, M2 (to the west-southwest of C1), and M3. 
The latter crossed C1 in 2010 Oct. (preliminary speed $\sim10\,c$\footnote{We assume along this paper the standard cosmology with $H_0$=71 km s$^{-1}$ Mpc$^{-1}$, $\Omega_M=0.27$, and $\Omega_\Lambda=0.73$.
Under this assumption the angular scale is 4.48\,pc/mas, and 1\,mas/yr  proper motion translates into a superluminal apparent speed of $19\,c$.}) and represents the clearer case of motion outwards C0 (Fig.~\ref{fig1}).
This is also evident from this movie: {\tt https://w3.iaa.es/$\sim$iagudo/research/OJ287/OJ287End2010.mov} , that starts just before the time of ejection of M3.

The most prominent 1\,mm flares ever reported in {OJ287} (${\rm{A}}_{\rm{mm}}$ and ${\rm{B}}_{\rm{mm}}$, see  Fig.~\ref{fig1}) took place in C1, as shown by the match of the 7\,mm light curve of C1 and those at other mm wavelengths.
From the half opening angle of the inner 7\,mm jet in {OJ287}  -- $1^\circ\kern-.35em .9$--$4^\circ\kern-.35em .1$ \cite{Jorstad:2005p264,Pushkarev:2009p9412} --, and the mean distance of C1 and C0 ($0.23\pm0.01$\,mas), we estimate that C1 is located $>14$\,pc downstream from C0 (i.e. the core).

\begin{figure}
   \centering
   \includegraphics[clip,width=12.cm]{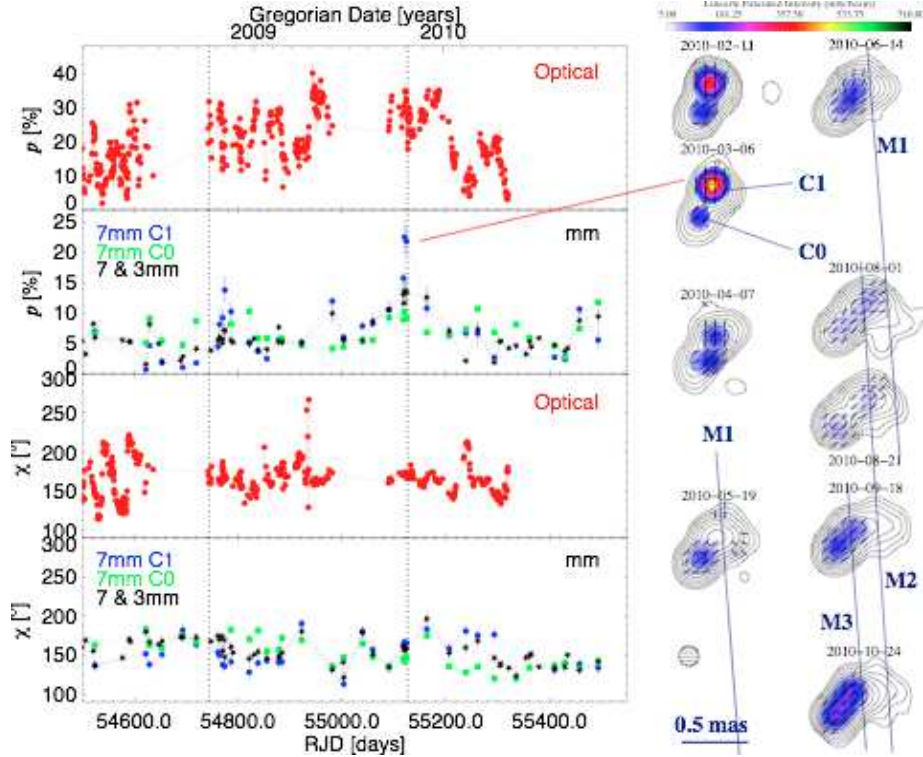}
   \caption{{\emph Left:} Optical and mm-wave linear polarization of OJ287 as a function of time.
   {\emph Right:} 7\,mm VLBA sequence of images as in Fig.~\ref{fig1}.}
   \label{fig2}
\end{figure}

\subsection{$\gamma$--ray flaring activity at the 7\,mm core}
The two ${\rm{A}}_{\rm{mm}}$ and ${\rm{B}}_{\rm{mm}}$ flares in {OJ287} start to rise at the time of two sharper $\gamma$--ray flares, as interpreted by \cite{Lahteenmaki:2003p5657} for a number of sources from the analysis of previous EGRET and Mets\"{a}hovi data.
These two flares, on 2008 October 4 and 2009 October 24, display $0.1$--$200$\,GeV photon fluxes a factor $\sim2$ and $\sim5$ larger than the quiescent $\gamma$--ray level of the source, respectively. 

\begin{figure}
   \includegraphics[clip,width=5.cm]{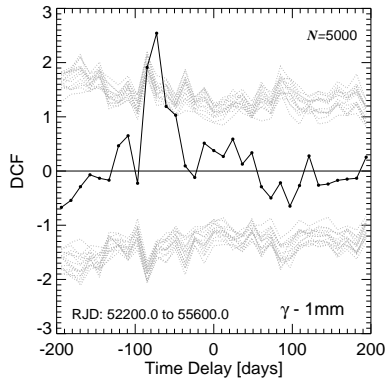}
   \begin{minipage}[b]
   {25pc}\caption{\label{fig3}DCF (as defined in \cite{Agudo:2011p14707}) between the $\gamma$-ray and 1\,mm light curves of OJ287. 
   Grey dotted curves of DCF values denote 99.7\% confidence limits for correlation (or anti-correlation at negative DCF) against stochastic variability as derived from Monte-Carlo simulations (see \cite{Agudo:2011p14707} for details about the method).}
   \end{minipage}
\end{figure}

The discrete correlation function (DCF) between the $\gamma$--ray to 1\,mm light curves (Fig.~\ref{fig3}), shows a prominent peak at a time lag $\sim-80$ days ($\gamma$--rays leading).
Our Monte-Carlo simulations to test the statistical significance of such correlation peak show that it is significant at 99.7\,\% confidence, hence confirming the correlation between ${\rm{B}}_{\gamma}$ and ${\rm{B}}_{\rm{mm}}$ --the two most luminous $\gamma$--ray and 1\,mm flares in our data.
Finally, the match between the C1 7\,mm light curve and the 1\,mm light curve locates the ${\rm{B}}_{\gamma}$ flare at the position of the C1 jet feature.

The optical light curve shows two sharp peaks at the peak time of ${\rm{A}}_{\gamma}$ and ${\rm{B}}_{\gamma}$.
However, the poorer X-ray time sampling does not allow us to claim for connection between the $\gamma$-ray and X-ray variability.

\subsection{Related linear polarization variability}
Despite the similarity of C0 and C1 in their $\chi$ evolution, these two emission regions differ in their $p$ behavior (Fig.~\ref{fig2}).
Whereas C0 never reaches $p\apgt10$\,\%, C1 exhibits the two largest $p$ peaks (of $14$\,\% around 2008--11--04, and $22$\,\% on $\sim$2009--10--16) ever observed in {OJ287} at 7\,mm.
These two violent $p$ increases of the core are contemporaneous with ${\rm{A}}_{\gamma}$ and ${\rm{B}}_{\gamma}$.
The mm $p$ peak in the end of 2008 seems to be delayed by $\sim15$ days with regard to ${\rm{A}}_{\gamma}$, while the second ($p\apgt22$\,\%) peak is consistent with being simultaneous with the extreme ${\rm{B}}_{\gamma}$ $\gamma$--ray peak.
We take this coincidence of these two exceptional events as evidence that both of them occurred at the same place: the C1 jet feature at $>14$\,pc from the central black hole.

The optical polarization fraction, which varies faster and stronger than that at mm wavelengths --consistent with the frequency dependence turbulence model of \cite{Marscher:2010p12402}--, also evidences sharp $p$ peaks at essentially the same time of the peak times of ${\rm{A}}_{\gamma}$ and ${\rm{B}}_{\gamma}$.
However, similar optical polarization fractions also happen at other times.

In contrast, the optical and mm linear polarization angle is surprisingly stable at $\chi\approx160^{\circ}$--$170^{\circ}$ (i.e. similar to the position angle of the inner jet) both during the time around ${\rm{A}}_{\gamma}$ and ${\rm{B}}_{\gamma}$ and in the overall $\chi$ evolution curves.
This stability is only greatly altered in the optical by sporadic short-term rotations of $\chi$ by up to $180^{\circ}$ \cite{Villforth:2010p11557}, consistent with turbulent plasma behavior.

\section{Discussion and Conclusions}

We report on a number of time coincidences in the multi--spectral--range and polarimetric behavior of {OJ287}.
These include:
{\it i)} two sharp 0.1--200\,GeV flares in October 2008 and 2009 (${\rm{A}}_{\gamma}$ and ${\rm{B}}_{\gamma}$, respectively),
{\it ii)} the rising of the two most luminous 1\,mm flares detected in the source (${\rm{A}}_{\rm{mm}}$ and ${\rm{B}}_{\rm{mm}}$), that our 7\,mm VLBA monitoring locates unambiguously at the position of C1,
{\it iii)} two sharp optical flares,
and {\it iv)} two sharp $p$ increases at the location of C1 --that are not accompanied by significant $\chi$ variability around a mean value $\bar{\chi}_{\rm{mm}}\approx160^{\circ}$--$170^{\circ}$.
Also, the extremely high linear polarization fraction of C1 during ${\rm{B}}_{\gamma}$ supports even stronger a location of the $\gamma$--ray events in C1.
 
The $\gamma$-ray inverse Compton (IC) emission is produced from either the synchrotron self-Compton (SSC) mechanism or IC scattering of infrared radiation from the dusty torus (IC/dust). 
An SSC model is certainly possible given the relatively small $\gamma$-ray to synchrotron luminosity ratio; $\approx 2$ in {OJ287}, see \cite{Agudo:2011p14707}. 
From the IC/dust mechanism it is expected that the optical and $\gamma$-ray emission should vary together \cite{Marscher:2010p11374}.
This is the case during the $\gamma$-ray flares, however IR emission from the dusty torus has not been detected so far in BL~Lacs like {OJ287}, which favors the SSC mechanism.

We propose a scenario where optical and $\gamma$-ray flares are produced by particle acceleration in C1 (interpreted as a conical standing shock \cite{1997ApJ_482L_33G,Agudo:2001p460}) when a moving jet feature interacts with C1, which is located well downstream the jet's acceleration and collimation zone (ACZ \cite{Marscher:2008p15675}), and at $>14$\,pc from C0 (i.e. the core).
The moving feature should be a turbulent perturbation with higher relativistic electron density and magnetic field than in the jet flow. 
The turbulent character is required to explain the weakness of the observed millimeter polarization outside the high polarization peaks. 
This scenario was treated by \cite{Cawthorne:2006p409}, whose model reproduces the polarization properties observed in OJ287 in the reported flares, i.e. large polarization peaks occurring at the time of interaction of turbulent blobs with a conical shock, and $\chi$ parallel to the jet axis (see \cite{Agudo:2011p14707}).
 
\begin{figure}
   \begin{center}
   \includegraphics[clip,width=15.cm]{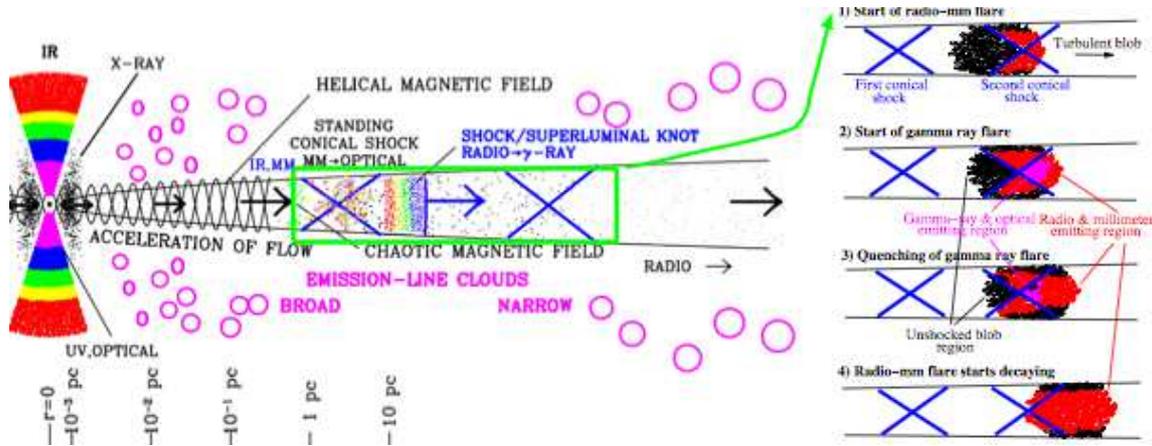}
   \caption{Illustration of the proposed scenario for the multi-spectral-range behavior of OJ287.}
   \label{fig4}
   \end{center}
\end{figure}

The mm and optical flares start with the increase of the magnetic field and electron energies of the moving feature after the interaction with the standing conical shock. 
The mm outburst continues as the lower-energy electrons fill the shocked region. 
The optical and $\gamma$-ray flares, produced by higher-energy electrons, are dumped near the conical shock. However, if synchrotron losses were the only mechanism producing the optical emission dumping in flares, the optical and $\gamma$-ray emission would be constant after the shocked jet feature fills the entire layer after the shock (at least before the entire jet feature crosses the conical shock). 
If the SSC mechanism dominates, the radiative losses are enhanced until the $\gamma$--ray losses start to be efficient, when electrons emitting optical synchrotron loose most of their energy and hence the optical flare is dumped.
The time-scale for this dumping is $(1+z)\delta^{-1} \sim 20 (\delta/19)^{-1} (a/0.054\,{\rm mas})$ days (where $a$ is the angular cross section of the jet and the Doppler factor ($\delta$) is the one reported by \cite{Jorstad:2005p264}), which is similar to the ${\rm{B}}_{\gamma}$ $\gamma$--ray time-scale; $\sim3$-$4$ weeks).

\ack
This research was partly funded by NASA grants NNX08AJ64G, NNX08AU02G, NNX08AV61G, NNX08AV65G, NNX08AW56S and NNX09AU10G, NSF grant AST-0907893, NRAO award GSSP07-0009; RFBR grant 09-02-00092; MICIIN grants AYA2007-67627-C03-03 and AYA2010-14844, and CEIC grant P09-FQM-4784.
The VLBA is an instrument of the NRAO, a facility of the NSF operated under cooperative agreement by AUI. 
The PRISM camera at Lowell Observatory was developed by Janes et al. 
The Calar Alto Observatory is jointly operated by MPIA and IAA-CSIC. 
The IRAM 30\,m Telescope is supported by INSU/CNRS, MPG, and IGN.
The SMA is a joint project between the SAO and the Academia Sinica. 

\section*{References}


\begin{thebibliography}{39}

\bibitem{Abdo:2010p11811} Abdo, A.~A., et~al. 2010, {\it Nature}, {\bf 463}, 919

\bibitem{Agudo:2001p460} Agudo, I., et~al. 2001, {\it ApJ}, {\bf 549}, L183

\bibitem{Agudo:2006p203} Agudo, I., et~al. 2006, {\it A\&A}, {\bf 456}, 117

\bibitem{Agudo:2010p12104} Agudo, I., Thum, C., Wiesemeyer, H.,  \& Krichbaum, T.~P.  2010, {\it ApJS}, {\bf 189}, 1

\bibitem{Agudo:2011p14707} Agudo, I., et~al. 2011a, {\it ApJ}, {\bf 726}, L13

\bibitem{Agudo:2011p15946} Agudo, I., et~al. 2011b, {\it ApJ}, {\bf 735}, L10

\bibitem{Cawthorne:2006p409} Cawthorne, T.~V. 2006, {\it MNRAS}, {\bf 367},  851

\bibitem{1997ApJ_482L_33G} G\'{o}mez et~al. 1997, {\it ApJ}, {\bf 482}, L33

\bibitem{Jorstad:2001p5655} Jorstad, S.~G.,  et~al. 2001a, {\it ApJS}, {\bf 134}, 181

\bibitem{2001ApJ...556..738J} Jorstad, S.~G.,  et~al. 2001b, {\it ApJ}, {\bf 556}, 738

\bibitem{Jorstad:2005p264} Jorstad, S.~G., et~al. 2005, {\it AJ}, {\bf 130}, 1418

\bibitem{Jorstad:2010p11830} Jorstad, S.~G., et~al. 2010, {\it ApJ}, {\bf 715}, 362

\bibitem{Lahteenmaki:2003p5657} L{\"a}hteenm{\"a}ki, A.,  \& Valtaoja, E. 2003, {\it ApJ}, {\bf 590}, 95

\bibitem{Marscher:2008p15675} Marscher, A. P., et~al, {\it Nature}, {\bf 452}, 966

\bibitem{Marscher:2010p12402} Marscher, A.~P.,  \& Jorstad, S.~G. 2010, {\it  Proc. Fermi meets Jansky --  AGN at Radio and Gamma-Rays}, p 171
  
\bibitem{Marscher:2010p11374} Marscher, A.~P., et~al. 2010, {\it ApJ}, {\bf 710}, L126

\bibitem{Pushkarev:2009p9412} Pushkarev, A.~B., Kovalev, Y.~Y., Lister, M.~L. \& Savolainen, T.  2009, {\it A\&A}, {\bf 507}, L33

\bibitem{Villforth:2010p11557} Villforth, C., et~al. 2010, {\it MNRAS}, {\bf 402}, 2087

\end{thebibliography}
\end{document}